\documentclass[10pt, a4paper]{article}
\usepackage{amssymb,amsmath,amsfonts}
\usepackage[dvipsnames]{xcolor}
\usepackage{graphicx}
\usepackage{longtable}
\usepackage{verbatim}
\usepackage{color}
\usepackage{mdframed}
\usepackage{soul}
\usepackage{mathrsfs}
\usepackage{amsfonts,amssymb,mathrsfs,amsmath,esint}    
\usepackage{slashed, cancel}
\usepackage{framed}
\usepackage{mdframed}
\usepackage{simplewick} 
\allowdisplaybreaks
\usepackage[bottom]{footmisc}
\usepackage{geometry}
\usepackage{framed}
\usepackage{mdframed}
\usepackage{tcolorbox}
\usepackage{latexsym}
\usepackage{graphicx}
\graphicspath{{./Figures/}}
\usepackage[dvipsnames]{xcolor}
\usepackage{booktabs}
\usepackage{datetime}
\newdateformat{mydate}{\THEDAY{ }\monthname[\THEMONTH]{ }\THEYEAR}

\usepackage{tikz}
\usepackage{color}
\usepackage{framed}
\usepackage{hyperref}
\hypersetup{colorlinks, citecolor=bluscuro, linkcolor=black, urlcolor=bluscuro}
\definecolor{rossos}{cmyk}{0,1,1,0.55}
\definecolor{bluscuro}{rgb}{0.15, 0.2, .85}
\definecolor{bluchiaro}{cmyk}{1,.3,0.,0.1}

\graphicspath{{./Figures/}}

\newcommand{\be}{\begin{equation}}
\newcommand{\ee}{\end{equation}}
\newcommand{\bea}{\begin{eqnarray}}
\newcommand{\eea}{\end{eqnarray}}
\newcommand{\beq}{\begin{equation}}
\newcommand{\eeq}{\end{equation}}
\newcommand{\lp}{\left(}
\newcommand{\rp}{\right)}
\newcommand{\llp}{\left[}
\newcommand{\rrp}{\right]}
\def\beqa{\begin{eqnarray}}

\def\pbh{{\text{\tiny PBH}}}

\def\d{{\rm d}}

\def\PBH{\text{\tiny PBH}}

\def\eeqa{\end{eqnarray}}

\def\lsim{\mathrel{\rlap{\lower4pt\hbox{\hskip0.5pt$\sim$}}
    \raise1pt\hbox{$<$}}}         
\def\gsim{\mathrel{\rlap{\lower4pt\hbox{\hskip0.5pt$\sim$}}
    \raise1pt\hbox{$>$}}}         

\def\lsim{~\rlap{$<$}{\lower 1.0ex\hbox{$\sim$}}}
\def\bsim{~\rlap{$>$}{\lower 1.0ex\hbox{$\sim$}}}

\def\ln{{\rm ln}}

\usepackage[normalem]{ulem}

\newcommand{\arXiv}[2]{\href{http://arxiv.org/pdf/#1}{{\tt [#2/#1]}}}
\newcommand{\arXivold}[1]{\href{http://arxiv.org/pdf/#1}{{\tt [#1]}}}

\numberwithin{equation}{section}
\renewcommand\theequation{\arabic{section}.\arabic{equation}}

\usepackage{amsmath}
\usepackage{amsfonts}
\usepackage{amssymb}
\usepackage{graphicx, rotating}
\usepackage{epstopdf}

\usepackage{epsfig}
\usepackage{latexsym}
\usepackage{graphicx}
\usepackage{color}
\usepackage{amsmath,amssymb}
\usepackage{cite}
\usepackage{slashed}

\usepackage{hyperref}
\hypersetup{colorlinks, citecolor=bluscuro, linkcolor=black, urlcolor=bluscuro}
\definecolor{rossos}{cmyk}{0,1,1,0.55}
\definecolor{bluscuro}{rgb}{0.15, 0.2, .85}
\definecolor{bluchiaro}{cmyk}{1,.3,0.,0.1}

\numberwithin{equation}{section}

\makeatletter
\def\section{\@startsection {section}{1}{\z@}{-3.5ex plus -1ex minus
-.2ex}{2.3ex plus .2ex}{\large\bf}}
\def\subsection{\@startsection{subsection}{2}{\z@}{-3.25ex plus -1ex
minus -.2ex}{1.5ex plus .2ex}{\normalsize\bf}}
\makeatother
\newcommand{\captionfonts}{\small}
\makeatletter  
\long\def\@makecaption#1#2{%
 \vskip\abovecaptionskip
 \sbox\@tempboxa{{\captionfonts #1: #2}}%
 \ifdim \wd\@tempboxa >\hsize
   {\captionfonts #1: #2\par}
 \else
   \hbox to\hsize{\hfil\box\@tempboxa\hfil}%
 \fi
 \vskip\belowcaptionskip}
\makeatother   

\catcode`@=11
\def\marginnote#1{}
\newcount\hour
\newcount\minute
\newtoks\amorpm
\hour=\time\divide\hour
by60
\minute=\time{\multiply\hour by60 \global\advance\minute
by-\hour}
\edef\standardtime{{\ifnum\hour<12 \global\amorpm={am}
\else\global\amorpm={pm}\advance\hour by-12 \fi
\ifnum\hour=0
\hour=12 \fi
\number\hour:\ifnum\minute<10
0\fi\number\minute\the\amorpm}}
\edef\militarytime{\number\hour:\ifnum\minute<10
0\fi\number\minute}
\def\draftlabel#1{{\@bsphack\if@filesw
{\let\thepage\relax
\xdef\@gtempa{\write\@auxout{\string
\newlabel{#1}{{\@currentlabel}{\thepage}}}}}\@gtempa
\if@nobreak
\ifvmode\nobreak\fi\fi\fi\@esphack}
\gdef\@eqnlabel{#1}}
\def\@eqnlabel{}
\def\@vacuum{}
\def\draftmarginnote#1{\marginpar{\raggedright\scriptsize\tt#1}}
\def\draft{\oddsidemargin
0.0truein
\def\@oddfoot{\sl preliminary draft \hfil
\rm\thepage\hfil\sl\today\quad\militarytime}
\let\@evenfoot\@oddfoot
\overfullrule 3pt
\let\label=\draftlabel
\let\marginnote=\draftmarginnote
\def\@eqnnum{(\theequation)\rlap{\kern\marginparsep\tt\@eqnlabel}
\global\let\@eqnlabel\@vacuum}
}
\catcode`@=12
\def\dj{\hbox{d\kern-0.347em \vrule width 0.3em height 1.252ex depth
-1.21ex \kern 0.051em}}

\def\d{{\rm d}}
\def\ee{{\rm e}\,}

\def\ba{\bar a}

\def\Dirac{\,\raise.15ex\hbox{/}\mkern-13.5mu D}
\def\dirac{\,\raise.15ex\hbox{/}\kern-.57em \partial}
\def\aslash{\,\raise.15ex\hbox{/}\mkern-13.5mu A}
\def\shalf{{\ifinner {\textstyle \frac{1}{2}}\else \frac{1}{2} \fi}}
\def\sthreehalfs{{\ifinner {\textstyle \frac{3}{2}}\else \frac{3}{2} \fi}}
\def\sshalf{{\ifinner {\scriptstyle \frac{1}{2}}\else \frac{1}{2} \fi}}
\def\sfourth{{\ifinner {\textstyle \frac{1}{4}}\else frac{1}{4} \fi}}
\def\sphifour{{\ifinner {\textstyle \frac{1}{4!}}\else \frac{1}{4!} \fi}}
\def\lsim{\stackrel{<}{_\sim}}

\def\XXint#1#2#3{{\setbox0=\hbox{$#1{#2#3}{\int}$}
    \vcenter{\hbox{$#2#3$}}\kern-.5\wd0}}

\def\bea{\begin{eqnarray}} \def\eea{\end{eqnarray}}
\def\be{\begin{eqnarray}} \def\ee{\end{eqnarray}} \def\nn{\nonumber}

\newcommand{\promille}{%
 \relax\ifmmode\promillezeichen
       \else\leavevmode\(\mathsurround=0pt\promillezeichen\)\fi}
\newcommand{\promillezeichen}{%
 \kern-.05em%
 \raise.5ex\hbox{\the\scriptfont0 0}%
 \kern-.15em/\kern-.15em%
 \lower.25ex\hbox{\the\scriptfont0 00}}

\def\d{{\rm d}}
\def\nn{\nonumber}

\def\cs2{c_{s}^{2}}

 \def\be   {\begin{equation}}   \def\ee   {\end{equation}}
 \def\ba   {\begin{array}}      \def\ea   {\end{array}}
 \def\bea  {\begin{eqnarray}}   \def\eea  {\end{eqnarray}}
 \def\bean {\begin{eqnarray*}}  \def\eean {\end{eqnarray*}}

\begin{document}
\parskip=10pt
\baselineskip=18pt

{\footnotesize  }
\vspace{5mm}
\vspace{0.5cm}

\begin{center}

\def\thefootnote{\fnsymbol{footnote}}

\topskip=70pt

{ \large 
\bf On the Primordial Black Hole Mass Function for Broad Spectra
}
\\[1.2cm]
{ V. De Luca, G. Franciolini  and A. Riotto} \\[0.6cm]

{\small \it D\'epartement de Physique Th\'eorique and Centre for Astroparticle Physics (CAP), \\
Universit\'e de Gen\`eve, 24 quai E. Ansermet, CH-1211 Geneva, Switzerland}

\vspace{.2cm}

\end{center}

\vspace{.8cm}

\begin{center}
\textbf{Abstract}
\end{center}
\noindent
We elaborate on the mass function of primordial black holes in the case in which the power spectrum of the curvature perturbation is broad. For the case of a broad and flat spectrum, we argue
that such a mass function is peaked at the smallest primordial black mass which can be formed and possesses a tail decaying like $M^{-3/2}$, where $M$ is the mass of the   primordial black hole. 
\vspace{.5in}

\def\thefootnote{\arabic{footnote}}
\setcounter{footnote}{0}
\pagestyle{empty}

{\let\thefootnote\relax \footnote{{\small Email:  Valerio.DeLuca@unige.ch, Gabriele.Franciolini@unige.ch, Antonio.Riotto@unige.ch}}}

\newpage
\pagestyle{plain}
\setcounter{page}{1}

\section{Introduction}
\noindent
Following the detection of  gravitational  waves sourced  by the  merging of two $\sim 30 M_\odot$ black holes \cite{ligo}, 
the idea that Primordial Black Holes (PBHs) might form   all (or a fraction of)  dark matter has recently  obtained  a lot of attention  (see Refs.~\cite{kam,rep1,rep2} and Ref.~\cite{revPBH} for a recent review). Of course, to answer the question about how much of the dark matter is composed of PBHs, one has to take into account
the  observational constraints on the PBH abundance  in a given mass range  \cite{Kuhnel:2017pwq,Carr16,co,Bellomo:2017zsr,Azhar:2018lzd,Lehmann:2018ejc}. This  requires accounting for the PBHs initial clustering \cite{ch1,ch2,Tada:2015noa,Franciolini:2018vbk,cl1,cl2,cl3,dizgah,Suyama:2019cst,Young:2019gfc} and other later phenomena as  merging and accretion.

PBHs may form in the early universe if there is an enhancement of the primordial curvature perturbation $\zeta$ generated during inflation \cite{s1,s2,s3} at  small-scales. Those small-scale perturbations are transferred to the radiation fluid after inflation through the re-heating process.
When such perturbations re-enter the cosmological horizon, if they are large enough in amplitude, they can collapse and form a PBH of mass similar to the horizon mass (see, for example, Refs.~\cite{gm,musco} for more details on the criterion for formation).

One of the basic quantity to robustly assess if    PBHs may or not form a given fraction of the dark matter is the so-called mass function $\psi (M)$, representing the fraction of PBHs with mass in $(M, M+ \d M)$, which is routinely defined by the relation \cite{revPBH}
\begin{equation}
\label{mf}
	\psi(M) = \frac{1}{\rho_\PBH} \frac{\d \rho_\PBH (M)}{\d M}
\end{equation}
and normalised such that
\begin{equation}
	\int  \psi(M)  \d M =1.
\end{equation}
While in the case in which the curvature perturbation power spectrum is peaked at a given wavelength the   mass function is peaked at the corresponding horizon mass when that given wavelength re-enters the horizon \cite{Niemeyer:1997mt,Yokoyama:1998xd,lg,Kuhnel:2015vtw}, in the case in which the curvature perturbation power spectrum is broad, the calculation of the mass function is far less intuitive.

In the present note  we elaborate on the mass function of PBHs for a broad power spectrum of the curvature perturbation. As a benchmark point,  we take the broad and  flat spectrum which is routinely  discussed  in the literature\footnote{For the production of   PBHs from large and small-scale curvature fluctuations in single field models of inflation,  a departure from slow-roll is needed \cite{hu}. A flat and broad power spectrum may be generated during a non-attractor phase when the inflaton potential is extremely flat. For those modes which exit the Hubble radius during such a phase
the corresponding power spectrum is flat as a result of a duality symmetry which maps the non-attractor phase into a slow-roll phase \cite{wands,biagetti}.} parametrised as
\be\label{eq:broad_specc}
{\cal P}_\zeta (k)\approx{\cal P}_0  \,  \Theta \left ( k_{s}-k\right) \Theta \left ( k-k_{l}\right),\quad k_s\gg k_l,
\ee
where $\Theta$ is the Heaviside step function\footnote{We will discuss the cases of a slightly   red- and blue-tilted broad spectra with a non-vanishing spectral tilt $n_p$ in the last section.}. We will call $M_s$ and $M_l$ the masses contained in the horizon when the wavelengths $\lambda_s\sim k_s^{-1}$ and $\lambda_l\sim k_l^{-1}$ re-enter the horizon, respectively.

The formation of the PBHs is a rare event. One should imagine our universe rescaled at a given time in the past containing
several Hubble volumes which are growing with time. At any instant of time there is a certain probability to form a PBH in each of these Hubble volumes. In the case of a broad power spectrum one expects that PBHs with different masses may form. This expectation is indeed correct and we argue that  
 the computation of the corresponding mass function can be performed based  on the  following rather simple physical  arguments:
\begin{itemize}
	\item[1)] the threshold for collapse is the same for perturbations entering the horizon (and collapsing) at different times;
	\item[2)] the PBHs formation probability is independent of the time of collapse. In other words, PBHs of different masses have the same formation probability; 
	\item[3)] the so called ``cloud-in-cloud'' problem of a PBH being absorbed by a bigger one  is irrelevant due to the low formation probability;
	\item[4)] the time evolution of the mass fraction privileges the smaller PBHs formed earlier.
	\end{itemize}
We will elaborate  these points one by one and show that the mass function has a peak at the mass $M_s$ corresponding to the horizon mass when the shortest wavelength $\lambda_s$
re-enters the horizon and a tail which scales like $M^{-3/2}$ for larger masses. Dynamically, what happens is that in each Hubble volume one may independently compute the probability of forming a PBH as a single event. This probability is the same for all masses and the final mass function has simply to account for the proper time evolution. 	

One should also recall that,  in the absence of some primordial non-Gaussianity,    PBHs  are not initially  clustered  even if the spectrum is broad \cite{dizgah} and therefore the mass function is not initially
modified by the clustering. 
Of course, it should be clear that we are not dealing here with the mass function obtained including much later and subsequent phenomena like merging and accretion.  One roughly expects that merging would push   the PBH mass distribution  towards larger masses from the  equality time to the current era.  Accretion would increase both the abundance and the mass of PBHs.  To the best of our knowledge, a thorough study is still missing and these issues will be solved most probably by running some dedicated N-body simulations (see Ref.~\cite{nbody} for a recent effort in this direction).

This note is organised as follows. In section 2 we deal with the threshold for collapse, in section 3 we show that the probability of formation  is the same for all PBHs masses. In section 4 we deal with the cloud-in-cloud problem while we discuss the mass function in section 5. Finally, we conclude and add some comments in section 6.

\section{The threshold for collapse is the same for all PBH masses}
\noindent
 One of the most important ingredients to characterise the probability of collapse is the critical threshold. It is understood that, in order to formulate a criterion for collapse, the overdensities peaks must be described in real space taking into account contributions from all Fourier modes.

 In linear approximation, the density contrast and the curvature perturbation are related as\footnote{In fact, the relation between the curvature perturbation and the density contrast is non-linear  and recently studied in Refs.  \cite{Yoo:2018kvb,ng1,ng2,ng3,ng4,Kalaja:2019uju,ng5}, but for the sake of simplicity of the arguments we neglect here its effects. We do not expect our findings to change considerably when non-linearities are taken into account. } 
 \begin{align}
	\frac{\delta \rho}{\rho_\text{\tiny b} } (k,\eta)&= \frac{4}{9} \lp\frac{k}{\cal H(\eta)} \rp^2 \zeta_{\vec k} (\eta) 
= \frac{4}{9} \lp\frac{k}{\cal H(\eta)} \rp^2  T(k \eta) \zeta_{\vec k}, 
\end{align}
where ${\cal H(\eta)}$ is the Hubble rate in conformal time and 
\begin{equation}
	T(k \eta) =3\frac{\sin (k \eta/\sqrt{3})-(k \eta /\sqrt{3})\cos(k \eta /\sqrt{3})}{(k \eta /\sqrt{3})^3}
\end{equation}
is the radiation  linear transfer function accounting for the time evolution of the curvature field, and effectively smoothing out subhorizon modes.
Assuming spherical symmetry, to determine the characteristic scale and amplitude of fluctuations that collapse to form a PBH, one defines the compaction function ${\cal C} (r)$ as the ratio of mass-excess within a sphere of radius $r$ to the areal radius at $r$ \cite{compac,musco}. The criterion for collapse widely used in the recent literature states that a PBH forms if the maximum of the compaction function is above a certain threshold. It can be also shown that the amplitude of the fluctuations in terms of the excess mass within a spherical volume is equivalent to the local value of the compaction function. Indeed, at a given conformal time $\eta$, one can compute the average profile of the overdensities as
 \be\label{av0}
\overline{\frac{\delta \rho}{\rho_\text{\tiny b} }} (r,\eta)=\delta_0\frac{\xi(r,\eta)}{\sigma^2(\eta)},
\ee
with $\xi(r,\eta)$ being the two-point correlator in real space if working in the comoving slicing, 
\be
\label{av}
\xi(r,\eta)=
\int\frac{{\rm d} k}{k}\frac{\sin k r}{k r}{\cal P}_{\delta \rho/\rho_\text{\tiny b}}(k,\eta),
\qquad 
	\sigma^2(\eta) = \xi(0,\eta)
\ee
and $\delta_0$ the amplitude of the perturbation.
The compaction function corresponding to the mean profile can be computed in terms of the volume averaged density contrast $\delta(r,\eta)$ using 
\begin{align}
	{\cal C}(r) &= r^2 {\cal H}^2 \delta (r,\eta) 
	\equiv r^2 {\cal H}^2 \llp \frac{3}{r^3} \int ^r_0 \d r\, r^2\, \frac{\overline {\delta \rho}}{\rho_\text{\tiny b} }(r,\eta)\rrp 
\end{align} 
which is time independent on super-Hubble scales.
The typical size of the perturbation in real space is given by the scale at which the maximum of the compaction function is located, which is usually denoted as $r_m(\eta)$.
Given that PBHs form from large-amplitude fluctuations, one can apply the peak statistics \cite{bbks}, assuming spherical peaks (which is a good approximation since PBHs are rare events), to calculate $r_m$. 
One can easily find that the size of the typical perturbation grows in time as $r_m(\eta) \sim \eta$ due to the smoothing of modes done by the transfer function.

The threshold for collapse $\delta_c(\alpha)$ is provided by numerical simulation and depends on the peaks profile through the parameter
	$\alpha = - ({\cal C}''(r_m) r_m^2/4 {\cal C}(r_m))$ \cite{musco,Escriva:2019phb}.
The dependence of the threshold on the parameter $\alpha$ is just a useful parametrisation of the shape dependence of the threshold. 

\begin{figure}[t!]
	\centering
	\includegraphics[width=.48\textwidth]{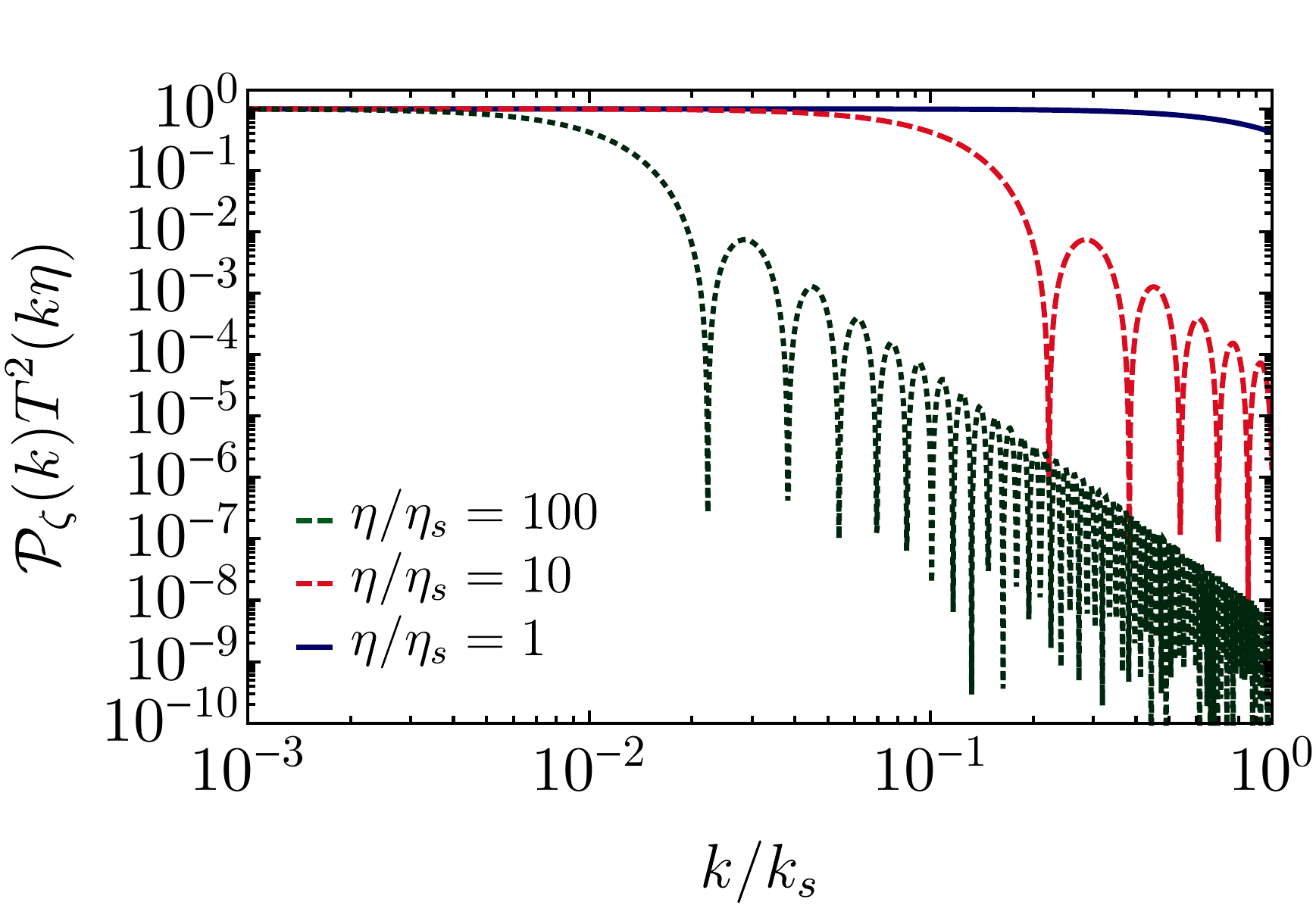}
	\includegraphics[width=.48\textwidth]{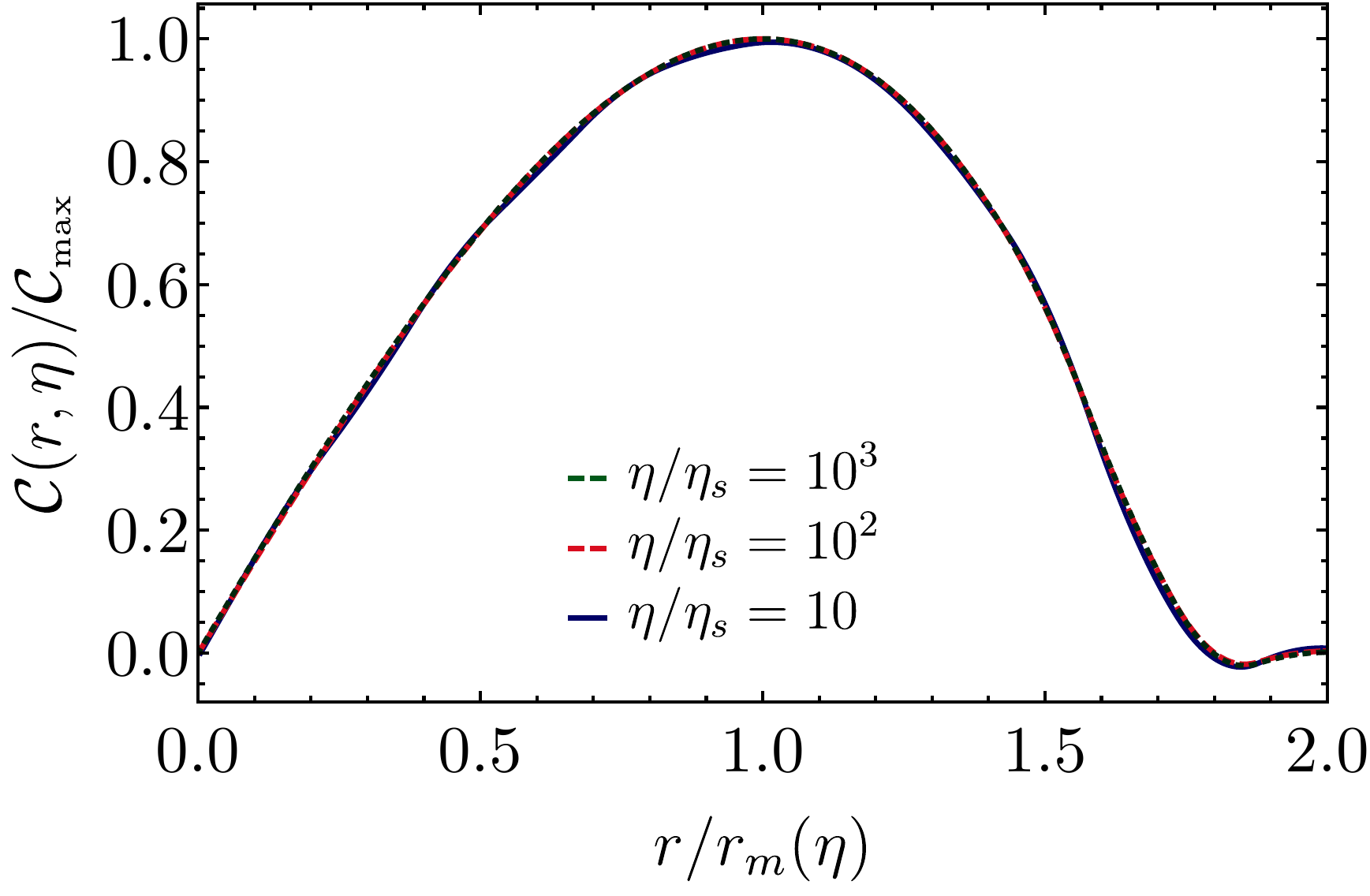}
		\caption{\it \textbf{Left:} Plot of the power spectrum of the comoving curvature perturbation. The transfer function effectively smooths out modes $k>1/\eta ={\cal H}$. That scale dominates in the computation of the profile giving characteristic peaks size  $r_m\sim 1/{\cal H}$.
\textbf{Right:} Rescaled compaction function depending on the time $\eta$. Each curve has been normalised to its corresponding $r_m(\eta)$.
}
			\label{fig1.1}
\end{figure}

Now, the crucial  point is that, at different times, the change in the mean profile is not enough to result in a significant change of  the threshold which is  $\delta_c\simeq 0.5$. This result holds for a tilted density power spectrum. This can be understood from first principles. In fact, a flat power spectrum of the comoving curvature perturbation corresponds to a blue tilted density  power spectrum ${\cal P}_{\delta \rho/\rho_\text{\tiny b}} (k, \eta) \sim k^4 \Theta({\cal H}(\eta)-k)$ at each time $\eta$, see Fig.~\ref{fig1.1}, where we also show that the rescaled compaction function for different times are basically the same. 
  Therefore, all mean profiles are close to $\overline{\delta \rho}/\rho_\text{\tiny b} (r, \eta)  \sim {\rm sinc}({\cal H(\eta)} r)$  which is the result for a narrow power spectrum for which $\delta_c= 0.5$, see also \cite{young}. Notice that   selecting different  wavenumbers  can be in principle done by a smoothing over a radius $\sim {\cal H}^{-1}$. In practice, we use the radiation transfer function to do this job. We do so also motivated by the fact that the evolution of the  perturbation should not depend upon the choice of the smoothing function as a matter of fact.

\section{The formation probability is the same for all PBH masses}
\begin{figure}[t!]
	\centering
	\includegraphics[width=.48\textwidth]{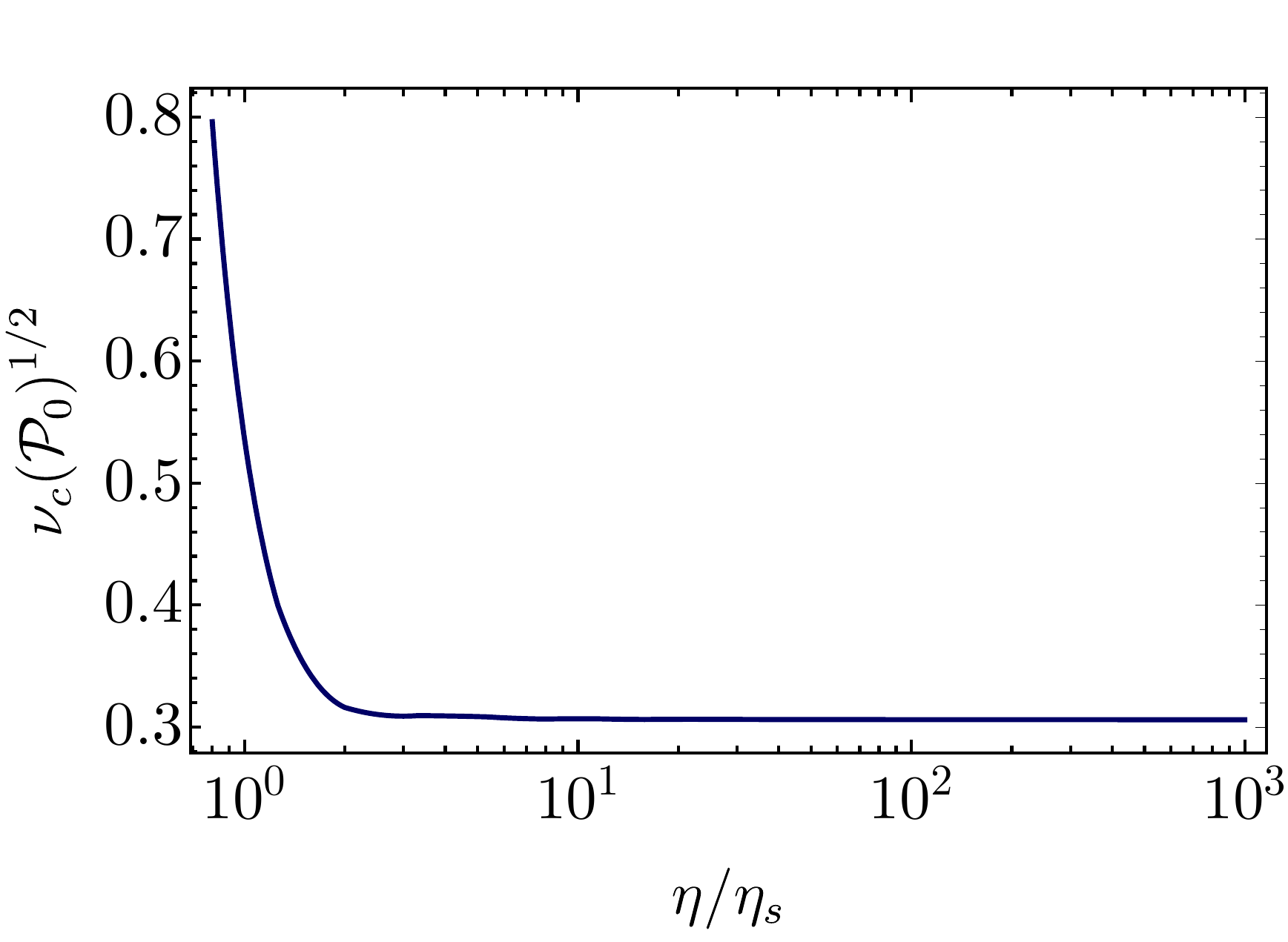}
		\includegraphics[width=.495\textwidth]{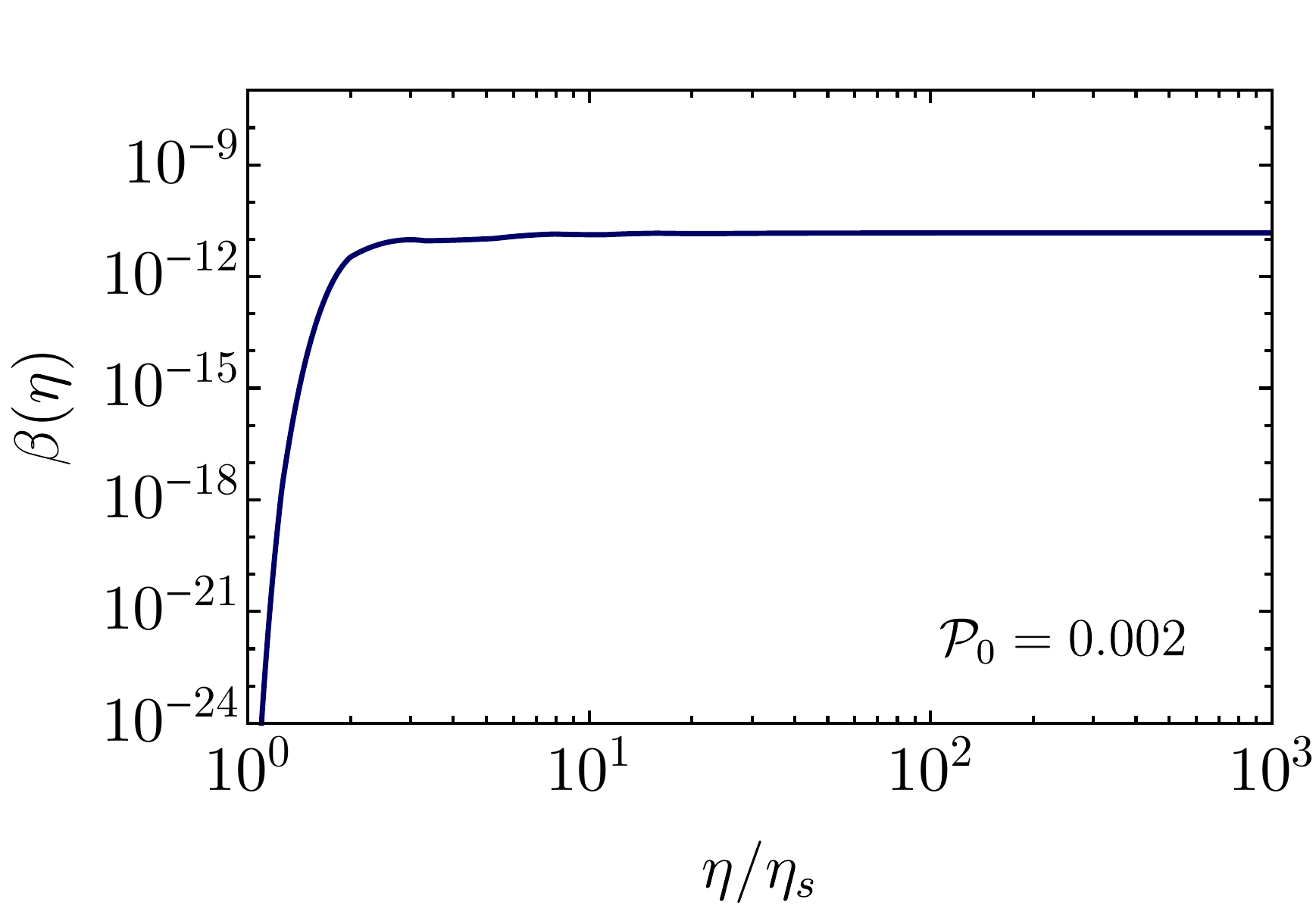}
		\caption{\it \textit{\textbf{Left:}}  The time dependence of the rescaled parameter $\nu_c$ as a function of time. The time $\eta_s$ identifies the horizon crossing of the smallest scale $\lambda_s$. For earlier times the combination scales like $\eta^{-2}$ due to the scaling of the variance $\sigma\sim 1/{\cal H}^2$ suppressing the formation probability for masses smaller than $M_s\equiv M(r_{m,s})$.
		\textit{\textbf{Right:}} Corresponding probability of formation $\beta(\eta)$ as a function of time for a reference amplitude ${\cal P}_0 = 0.002$. 
		} 
		\label{fig1}
\end{figure}

\noindent
Let us sketch in this section the procedure to compute the probability of forming a PBH of mass $M(r_m(\eta))$. 
A black hole forms if $\delta ( r_m,\eta) \geq \delta_c$, and once the critical threshold is known, 
the abundance of PBHs can be computed using the Press-Schechter formalism \cite{PS}, assuming Gaussian probability distribution for the curvature perturbation, as
\begin{align}
\beta(r_m(\eta))&=\int_ {\delta_c}  \frac{\d \delta}{\sqrt{2 \pi \sigma^2}} {\rm e} ^{-\delta ^2 / 2 \sigma^2}
= \frac{1}{2} {\rm erfc} \llp \frac{\nu_c}{\sqrt{2}}\rrp,
\end{align}
where $\nu_c \equiv \delta_c/ \sigma$. For completeness we report the full expression of the variance $\sigma$ which is
\begin{equation}\label{var}
	\sigma^2(\eta) = \frac{16}{81} \int \frac{\d k}{k} \frac{k^4}{{\cal H}^4}
	{\cal P}_\zeta (k) T^2(k \eta) W^2 (k, r_m(\eta))
\end{equation}
with 
$r_m(\eta)\simeq 3 /{\cal H}(\eta)$
 and 
\begin{equation}
W(k,r_m(\eta))=3\frac{\sin (kr_m)-(kr_m)\cos(kr_m)}{(kr_m)^3}.
\end{equation}
One finds a constant $
	\nu_c ({\cal P}_0)^{1/2} \sim 0.3$,  see Fig.~\ref{fig1}. 
 This is the result of two ingredients. The first is  the time independence of the threshold while the second is the effect given by the horizon growth which counteracts the action of the transfer function smoothing out subhorizon modes  in  Eq.~\eqref{var}. 
For a slightly red (blue) tilted power spectrum, the exponential dependence of the probability to $\nu_c$ implies that the larger (smaller) PBHs are much more likely to form. 

There exists an additional property of the PBHs formation that needs to be taken into account when computing the probability of formation of a PBH of mass $M$ at a given epoch. Indeed,  it was found that the masses of the PBHs respect a scaling relation near the critical threshold for collapse $\delta_c$ as \cite{Evans:1994pj,Niemeyer:1997mt,Choptuik:1992jv}
\begin{equation}
\label{Mcri}
M (\delta)= {\cal K} M_H \lp \delta - \delta_c\rp^\gamma
\end{equation}
where we take ${\cal K} =4$ as a reference value and $\gamma=0.36$ in a radiation dominated universe \cite{koike,Musco:2004ak,Musco:2008hv,Musco:2012au,Escriva:2019nsa}.
The mass function of PBHs produced at a given epoch, see for example \cite{Niemeyer:1997mt}, possesses a low-mass  tail scaling like $\sim M^{2.8}$ and an exponential fall-off after the peak located around 
$M \sim M_H$. This effect dominates the low-mass portion of the final mass function while, since the distribution of masses produced at a given epoch is  fairly peaked, this scaling relation does not play any major role in shaping the high-mass range of the final mass fraction $\psi (M)$, as we will see in the following.

\section{Lighter PBHs are not absorbed by heavier ones}

One of the key issues in determining the PBH mass function is that, in each of the horizon volumes, the probability that a PBH is formed and subsequently absorbed by  a bigger and more massive one is totally negligible. This point  is rather intuitive as the formation of a PBH is an extremely rare event, having two (or more) PBHs forming on the same site is even more improbable. To assess this point we can use the so-called excursion set method \cite{Bond}. In this method the density perturbation performs a random walk  as a function of the   smoothing scale, and the PBH  formation problem becomes the  so-called first-passage time problem in the presence of a barrier. The cloud-in-cloud problem is thus taken care of by considering  only those trajectories which cross  the threshold for the first-time.
The results of this section are not original and contained in Ref. \cite{dizgah}, but we summarise them here for the reader's benefit.

As in the previous section, let us define the  smoothed density contrast 
\be\label{dfilter}
\delta({\vec x},R) =\int \d^3x'\,  W(|{\vec x}-{\vec x}'|,R)\, \delta({\vec x}'),
\qquad 
\delta(\vec k,R) = \widetilde{W}(k,R) \delta(\vec k),
\ee
where, in this section,  $\delta({\vec x})=\delta \rho/\rho_\text{\tiny b} ({\vec x}) $,  $W(|{\vec x}-{\vec x}'|,R)$ is the window function and $R$ identifies the particular smoothing scale. 
We also employed the fact that the convolution in real space becomes a product in momentum space.
A particularly convenient choice for the window function is provided by a sharp filter in $k$-space as\footnote{The reader might at this point be confused by the fact that we choose a window function different from the one in the previous section. This is motivated by the fact that choosing  a step function in real space would cause  the random walk to be slightly   non-markovian  \cite{MR1,MR2,MR3}. We do not expect the results discussed in this section  to be affected by it.}
\be\label{Wsharpk}
\widetilde{W}_{{\rm sharp}-k}(k,k_\text{\tiny f})=\Theta(k_\text{\tiny f}-k),
\ee
with $k_\text{\tiny f}=1/R$. As we will appreciate in the following, with such a choice, the excursion set theory simplifies considerably.
Using the Press-Schechter formula for the probability of collapse, one easily notices that such a probability depends only on the ratio  $\delta_c/\sigma(R)$, with $\sigma (R)$ being the variance of the smoothed field.
Therefore, in the excursion set scheme, it is useful to factorise the time evolution of the density contrast and absorb it in the threshold which conversely becomes a time dependent quantity denoted $\omega(a)$, which will be defined in the following. Thus, the density field in Eq.~\eqref{dfilter} becomes independent of time.
  We will discuss how this can be done in practice in the PBH scenario at the end of this section.

One then studies the ``evolution" of the density contrast  $\delta({\vec x},R)$ with respect to the smoothing scale  $R$, in a given position  ${\vec x}$, by defining 
\be\label{padeta1}
\frac{\partial\delta(R)}{\partial R} \equiv \kappa (R)=
 \int \frac{\d^3k}{(2\pi)^3}\,
\widetilde{\delta}({\vec k} ) \frac{\partial\widetilde{W}(k,R)}{\partial R}.
\ee
where we set the coordinates such that $\vec x = 0$ without loosing generality and designate  $\delta(R)$ as the smoothed density contrast field at the origin.
$\widetilde{\delta}({\vec k} )$ is, by construction, a stochastic variable, and therefore also $\kappa(R)$ inherits this property.
Then Eq.~(\ref{padeta1}) takes the form  of a Langevin equation, with $R$ being the  time variable and
$\kappa(R)$ being the stochastic noise.
The two-point connected correlator fully describes the properties of the Gaussian stochastic fields $\delta(R)$ and $\kappa(R)$. The two-point function of 
$\kappa(R)$ is 
\be\label{corr2eta}
\langle \kappa(R_1)\kappa(R_2)\rangle =
\int_{-\infty}^{\infty}\d  \, \ln k\,\, \Delta^2(k)
\frac{\partial\widetilde{W}(k,R_1)}{\partial R_1}
\frac{\partial\widetilde{W}(k,R_2)}{\partial R_2},
\ee
where $\Delta^2(k)=k^3P_\delta(k)/(2\pi^2)$.  The r.h.s. is a generic function of the smoothing scales $R_1$ and $R_2$.

However, in the case of the filter being a top hat in momentum space, one achieves significant simplifications. If one defines $Q (k_\text{\tiny f})=-(1/k_\text{\tiny f}) \kappa(k_\text{\tiny f})$ and recalling the definition $k_\text{\tiny f}=1/R$, one finds
\be
\label{padeta2}
\frac{\partial\delta(k_\text{\tiny f})}{\partial \,\ln \,k_\text{\tiny f}} =Q (k_\text{\tiny f})
\qquad \text{with} \qquad 
\langle Q (k_{\text{\tiny f}_1})Q (k_{\text{\tiny f}_2})\rangle =
\Delta^2(k_{\text{\tiny f}_1}) \delta_D(\ln\, k_{\text{\tiny f}_1}-\ln\, k_{\text{\tiny f}_2}).
\ee
Additionally, the variance of the time independent overdensities $S$ can be employed as ``time'' variable
\be\label{mu2RW2}
S(R) =\int_{-\infty}^{\infty} \d \,\ln\, k \, \Delta^2(k) \, |\widetilde{W}(k,R)|^2=
\int_{-\infty}^{\ln \,k_\text{\tiny f}} \d\,\ln\, k\, \Delta^2(k),
\ee
where, in the second step, we introduced our choice of sharp filter in momentum space as in Eq.~\eqref{Wsharpk}.
The equation of motion of the rescaled variable  $\xi (k_\text{\tiny f}) =Q(k_\text{\tiny f})/\Delta^2(k_\text{\tiny f})$ becomes 
\be\label{Langevin1}
\frac{\partial\delta(S)}{\partial S} = \xi(S),
\ee
where the stochastic noise is characterised by the two-point function
\be\label{Langevin2}
\langle \xi(S_1)\xi(S_2)\rangle =\delta_D (S_1-S_2).
\ee
Therefore, a Langevin equation with a Gaussian white noise describes the evolution of the density field as a function of the variance.
As a consequence, the evolution of $\delta(S)$ can be regarded as a Brownian random walk, with respect to $S$ which can be considered a ``time'' variable, with no memory effects being present.
As in Ref. \cite{Bond},  the particular realisation of the stochastic evolution of $\delta (S)$ will be referred to as a ``trajectory'' (see \cite{dizgah} and refereces therein for a schematic picture of this evolution).

It is a standard result that the distribution solving the Fokker-Planck equation
\be\label{FPdS}
\frac{\partial P}{\partial S}=\frac{1}{2}\, \frac{\partial^2 P}{\partial \delta^2}.
\ee
describes the stochastic variable which evolves as dictated by the Langevin equation in Eq.~\eqref{Langevin1}.

Now we need to insert notion of a time dependent threshold (or barrier in the brownian motion interpretation of the problem) in the picture. As described above, the time dependence of the density contrast is re-absorbed into a time dependent threshold for collapse.
In the case of PBHs formation, the time dependent threshold can be defined as $\omega(a) \equiv \delta_c/a^2$, then the barriers at two different times can be found using the relation
\be\label{rt}
\omega(a_1) = \left(\frac{a_2}{a_1}\right)^2 \omega(a_2).
\ee
Said in other terms, the barrier decreases with the passage of time while, as specified above, the variance stays constant. 
Therefore, the evolution with respect to the real time is captured by the time evolution of the barrier corresponding to the evolution of the comoving Hubble length.

For PBHs, similarly to the case of  halos, each trajectory provides the variation of the density contrast $\delta(R)$ as a function of the smoothing scale at a given time. 
However, a PBH forms if the density contrast is above threshold when that particular scale $\sim R$ crosses the horizon. Therefore, 
in order to compute the probability of collapse at the horizon crossing time, one needs to compute trajectories at a fixed time and 
evolve the values of the smoothed density contrast backwards or forwards in time until the scale of interest crosses the horizon, which is
$R\sim {\cal H}^{-1}$ for a PBH of mass $M_H$ (i.e. the mass contained within the Hubble volume),  which scales like $R \sim a\sim M_H^{1/2}$ in a radiation-dominated universe. 

For a broad and flat spectrum, see Eq. (\ref{eq:broad_specc}),  and still at fixed time, 
\be
\label{aa}
S(R_1)
=S(R_2) \left(\frac{M_{H_1}}{M_{H_2}}\right)^{-1}.
\ee 
The hierarchy for the thresholds and the variances is then given by
\be
\omega(a_s)\gg \omega(a_l)\quad {\rm and} \quad S_s\gg S_l.
\ee
However, one has that
\be
\sqrt{\frac{S_s}{S_l}}=  \frac{\omega(a_s)}{\omega(a_l)}.
\ee
This condition implies and confirms that small PBHs are generated with the same abundance as large ones. 

The ``two-barriers'' problem \cite{LC}  permits the computation of the probability to form a virialized object or a PBHs,  giving also the conditional probability that a particular trajectory, with an initial  up-crossing of the barrier $\omega_1$ at a certain value $S_1$ of the variance, will have a first up-crossing of $\omega_2$ between $S_2$ and $S_2+\d S_2$ with $S_1\gg S_2$ and $\omega_1>\omega_2$. This is directly related to the cloud-in cloud problem. 
The probability that a large PBH incorporates an already formed smaller PBH with $S(<S_s)$ at later times is given by \cite{LC} 
\begin{align}
P(S,\omega(a)|S_s,\omega(a_s))&= 
\frac{1}{2} {\rm Erfc} \lp \frac{\omega(a)}{\sqrt{2 S}} \rp 
\llp 1+ \exp \lp 2 \frac{ \omega (a)}{\omega (a_s)} \frac{\omega ^2(a_s)}{S_s}\rp \rrp
\simeq \beta(M(r_{m}))\ll 1,
\label{h}
\end{align}
where we employed $\omega(a_s)/\sqrt{S_s} = {\cal O}(6 \div 8)$ in the typical PBHs scenario, and the large enhancement which could be given by the exponential factor is teamed by the small coefficient $\omega(a)/\omega(a_s) \ll 1$.

This result can be understood intuitively by realising that, since the formation of a PBH is a rare event, the probability that a small  PBH is eaten by a larger one is very small, scaling like 
$\beta(M(r_{m}))\beta(M_s)\simeq \beta^2(M_s)$ for a flat broad spectrum \cite{dizgah}. One can wonder what happens if the mass hierarchy between $M_s$ and $M(r_m)$ is not so pronounced. In such a case, the expression (\ref{h}) may not be applied. However, one has to recall that the collapse starting with wavelength $\lambda_s$ entering the horizon lasts for a few Hubble times. This implies that the modes which enter immediately after $\lambda_s$ are involved in such a local collapse and therefore, in practice, the dynamics separates the masses and generates an effective hierarchy. This argument is  confirmed by numerical simulations \cite{gm,musco}.

\section{The PBH mass function for broad and flat spectra}

Having established the absence of the cloud-in-cloud problem for broad flat spectra, and since the abundance $\beta$ is the same for all masses, we are ready to compute the PBH 
 mass distribution. We recall that 
\be
R\sim k^{-1}\sim  1/aH\sim a\sim M_H^{1/2}.
\ee
Taking into account that the density of small-mass PBHs grows like non-relativistic matter, and that $\beta=\rho_\pbh/\rho_{\rm tot}$ at the time of the PBH formation, the abundance of the small PBHs at the formation time $\eta_l$ of the large ones is \cite{dizgah}
\begin{align}\label{5.2}
	\frac{\rho_s (\eta_l)}{\rho_l (\eta_l)} &= \frac{\rho_s (\eta_s) (a_s/a_l)^3}{\rho_l (\eta_l)}
	=
	\frac{\rho_s (\eta_s) (a_s/a_l)^3}{\rho_l (\eta_l)} \frac{\rho_{\rm tot} (\eta_l) a_l^4}{\rho_{\rm tot} (\eta_s) a_s^4}
	\nn \\
	&=\frac{\beta(M_s)}{\beta(M_l)} \frac{a_l}{a_s}
	=\frac{\beta(M_s)}{\beta(M_l)} \frac{k_s}{k_l}
	\nn \\
	&= \frac{k_s}{k_l}=\frac{M_l^{1/2}}{M_s^{1/2}}\gg 1.
\end{align}
Therefore, even though the probability of forming a PBH is independent of the mass, i.e. $\beta(M_s)=\beta(M_l)$, the PBHs with the smallest  mass will give the largest contribution to the mass distribution. 
 \begin{figure}[t!]
	\centering
	\includegraphics[width=.48\textwidth]{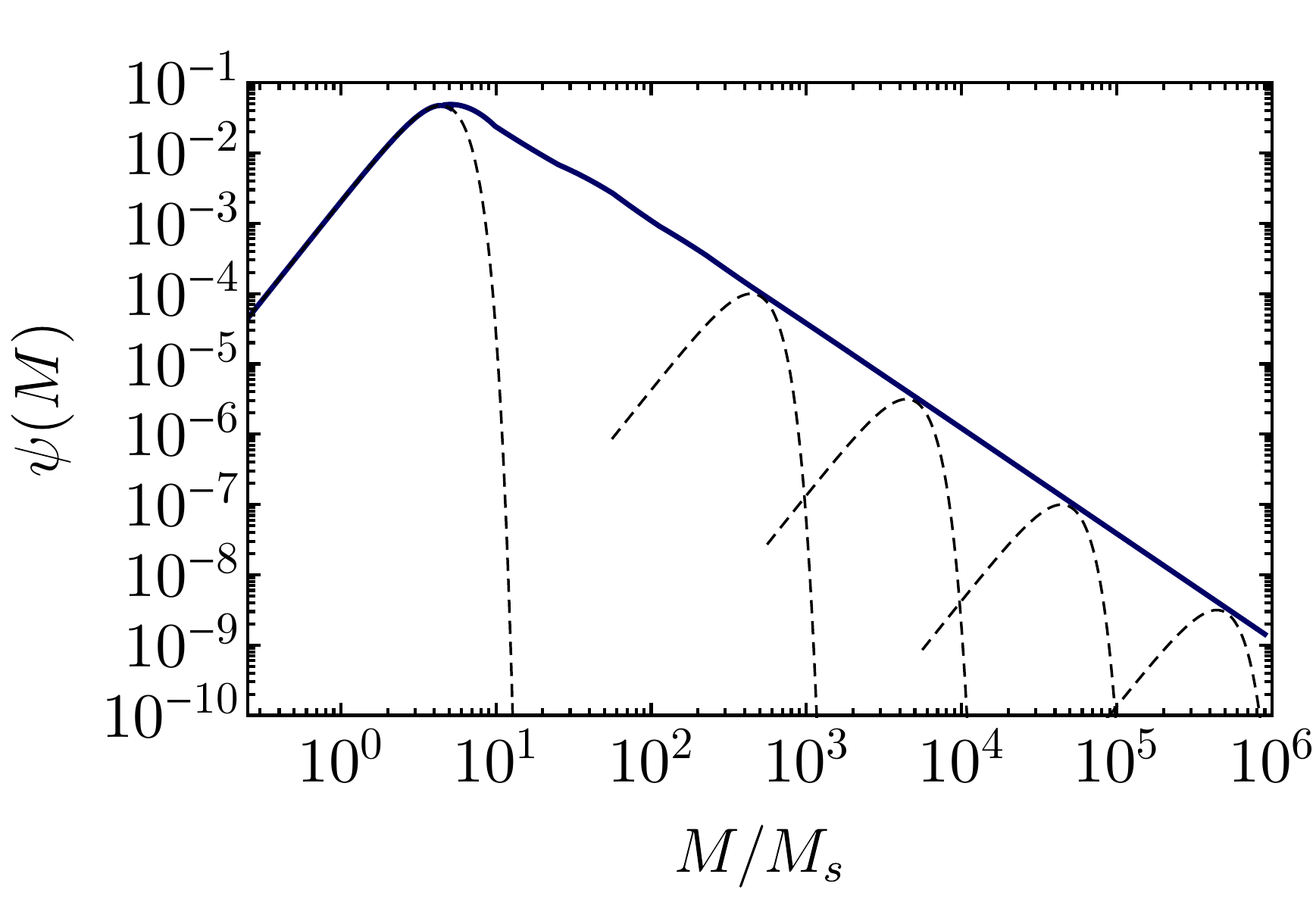}
				\caption{\it The PBH mass function for a broad and flat power spectrum of the curvature perturbation.
				The dashed black lines indicate the PBHs formations at different times which gives rise to the final $\psi(M)$ as their envelope.
				The low-mass tail is a feature given by the critical collapse at the time of re-entry of the smallest scale $\lambda_s$.
				} 
		\label{fig3}
\end{figure}
The resulting mass function can be then written as
\begin{tcolorbox}[colframe=white,arc=0pt]
		\vspace{-.4cm}
\begin{align}
	\psi(M)  &=  
		\frac{1}{f_\PBH}
		\frac{a_\text{\tiny eq}}{a_\text{\tiny f}} \frac{ \beta(M_H)}{M} 
	=
	\frac{1}{f_\PBH}
	 \frac{M_\text{\tiny eq}^{1/2}}{M_H^{1/2}}  \frac{ \beta(M_H)}{M} 
	 \simeq
	\frac{1}{f_\PBH}
	 \frac{M_\text{\tiny eq}^{1/2}}{M ^{3/2}} \beta(M),
\end{align}
\end{tcolorbox}
\noindent
where $M_\text{\tiny eq} \simeq 2.8 \cdot 10^{17} M_\odot$ is the horizon mass at the time of equality, $a_\text{\tiny f}$ indicates the scale factor at formation and from Eq.~\eqref{Mcri} one can use the approximate relation $M \simeq 0.9 M_H$. As we argued above, for the  flat power spectrum the probability $\beta(M)$  is constant, thus one gets a $M^{-3/2}$ tail for large masses, see also Refs. \cite{ds93,qcd,Byrnes:2018txb}.
Notice that the same result is found by using the time evolution found in \eqref{5.2}. Indeed, expanding for a mass $M+\d M$ one finds
\begin{equation}
\frac{\rho_s (M+\d M )}{\rho_l  (M_\text{\tiny eq})} -\frac{\rho_s (M)}{\rho_l  (M_\text{\tiny eq})} = \lp \frac{M_\text{\tiny eq}}{M+\d M }\rp^{1/2} -\lp \frac{M_\text{\tiny eq}}{M }\rp^{1/2} 
\sim - \frac{M_\text{\tiny eq}^{1/2}}{M ^{3/2}} \d M.
\end{equation}
Fig.~\ref{fig3} shows the  PBH mass function for a broad power spectrum of the curvature perturbation in relation to the PBH mass $M_s$ related to the smallest scale $\lambda_s$.

Notice also that, had one used a different definition of the mass function
\be
f(M)=f_\PBH\, M\, \psi(M) \,\,\,\,\,\,{\rm or}\,\,\, \,\,\,\int \d\,\ln\, M\, f(M)=f_\PBH,
\ee
the fall-off of the such mass function is less pronounced and maybe more directly related to the  observational constraints as a function of the PBH mass.

\section{Comments and conclusions}
In this short note we have calculated the PBH mass function for the case of a flat broad power spectrum of the curvature perturbation. We have argued that the mass function should be peaked at the mass corresponding to the wavelength entering the horizon first, i.e. the smallest PBH mass, with a tail decaying as $M^{-3/2}$. This finding originates from realising that PBHs with different masses have the same probability to form, that there is no cloud-in-cloud problem and that the tail is originated by the time evolution.

Let us close with some comments. One can ask what might happen if the power spectrum is broad, but slightly  red- or blue-tilted. Consider for instance
a broad power spectrum of the form
\be\label{eq:broad_speccc}
{\cal P}_\zeta \approx{\cal P}_0  \,(k/k_s)^{n_p}  \Theta \left ( k_{s}-k\right) \Theta \left ( k-k_{l}\right),\quad k_s\gg k_l,
\ee
where positive (negative) values of the index $n_p$ give rise to a blue (red) spectrum. First of all, in both cases, the cloud-in-cloud problem is absent \cite{dizgah}. Second of all, we find that the threshold satisfies the relation $
	\nu_c ({\cal P}_0)^{1/2} \sim 0.3 (\eta/\eta_s)^{n_p/2}$. It is rather intuitive to understand that in the case of a blue spectrum, the  generation of PBHs will be dominated by the smallest mass possible $M_s$. 
One might think  that when wavelengths $\gg \lambda_s$ re-enter the horizon, PBHs  are generated provided that  the corresponding variance is  sufficiently sizeable. This would imply however that the variance of the density contrast at the shortest scale  $\lambda_s$ would be much larger than unity and therefore outside of the perturbative regime. We do expect therefore that  broad spectra with a blue tilt predict  a mass function similar to those generated by narrow spectra. For a red spectrum, the variances computed at  horizon crossing for the longer wavelengths are  bigger than those for the  small-scale  perturbations. The generation  of PBHs is therefore dominated by the largest possible mass in the power spectrum $M_l$. 
One could think  that the generation of   the low mass  PBHs could be increased at later times due to the influence of the longer modes perturbations on the small ones after they have entered the horizon \cite{cl3}. However, such longer modes  re-enter the horizon when the short ones have have already done so by some time and the  radiation pressure has already caused the dispersion of the   peaks at small scales. This effect is described by  the radiation transfer function which decays as  $1/(k\eta)^2\ll 1$ on subhorizon scales.
One could also argue that  PBHs might be also  generated at small-scales. However, this  will require the hierarchy $\sigma_s \gg \sigma_l$ and therefore the  non-linear regime would have been achieved much earlier  the long mode has re-entered the horizon, thus making  all the computations again not reliable. 
It would be interesting to investigate the impact of the mass function we have derived onto the PBH constraints and the implications, e.g., for the LIGO events, see for instance Ref.  \cite{bp}.  

Finally, let us draw the reader's attention that recently a discussion of the mass function has appeared in Refs.  \cite{jap,ng5}. Their approach is similar and inspired by a peak theory-like approach. In particular, their condition that the compaction function has a maximum at a given smoothing scale $R$ should correspond to the first-time passage requirement in the excursion set. The latter  has also the advantage of  making a well-defined prediction about the cloud-in-cloud problem. 
Furthermore, while Refs.  \cite{jap,ng5} do not have an explicit expression for the mass function for a broad spectrum, we expect our results, based on simple physical arguments,  to be reproducible by their formalism once the  time evolution is accounted for in  the linear variances they adopted. 

\setcounter{section}{5}
\noindent

\begin{center}
{\bf  Acknowledgments}
\end{center}
\noindent
We thank C. Byrnes and  I. Musco for reading a version of the draft and for useful discussions. We also thank Y. Dalianis for providing us with a useful reference. The authors  are  supported by the Swiss National Science Foundation (SNSF), project {\sl The Non-Gaussian Universe and Cosmological Symmetries}, project number: 200020-178787.



\end{document}